\documentclass[reprint,aps,pra,showpacs]{revtex4-1}
\usepackage{graphicx}
\usepackage{dcolumn}
\usepackage{float}
\usepackage{hyperref}
\usepackage{latexsym}
\usepackage{amssymb}
\usepackage{amsfonts}
\usepackage{amsmath}
\usepackage{fancybox}
\usepackage[binary,squaren]{SIunits}
\usepackage{natbib,graphicx,dcolumn,color,lipsum}
\usepackage{subfigure,amsmath,amssymb,verbatim,moreverb,bm,framed}
\usepackage[sort&compress]{natbib}
\usepackage{colordvi}

\def\be{\begin{equation}}
\def\ee{\end{equation}}
\def\ber{\begin{eqnarray}}
\def\eer{\end{eqnarray}}

\def\sigmabold{\mbox{\boldmath $\sigma$}}
\def\rv{{\bf r}}

\def\pv{{\bf p}}

\def\nn{\nonumber}
\def\nn{\nonumber}

\def\be{\begin{equation}}
\def\ee{\end{equation}}
\def\ber{\begin{eqnarray}}
\def\eer{\end{eqnarray}}
\def\nn{\nonumber}
\newcommand{\agrave}{\`a}

\begin{document}
\title{The Edelstein effect in the presence of  impurity spin-orbit scattering}

\author{Amin Maleki}
\email{corresponding author;
	e-mail: maleki@fis.uniroma3.it}
\affiliation{Dipartimento di Matematica e Fisica, Universit{\agrave} Roma Tre, Via della Vasca Navale 84, 00146 Roma, Italy }

\author{Ka Shen}
\affiliation{Kavli Institute of NanoScience, Delft University of Technology, Lorentzweg 1, 2628 CJ Delft, The Netherlands}
\author{Roberto Raimondi}
\affiliation{Dipartimento di Matematica e Fisica, Universit{\agrave} Roma Tre, Via della Vasca Navale 84, 00146 Roma, Italy }
%
%



\begin{abstract}
In this paper we study the current-induced spin polarization in a two-dimensional electron gas, known also as the Edelstein effect. Compared to previous treatments, we consider both the Rashba and Dresselhaus spin-orbit interaction as well as the spin-orbit interaction from impurity scattering. In evaluating the Kubo formula for the spin polarization response to an applied electric field, we explicitly take into account the side-jump and skew-scattering effects.  We  show that the inclusion of side-jump and skew-scattering modifies the expression of the current-induced spin polarization.
\begin{description}
\item[PACS] 72.25.-b, 71.70.Ej, 72.20.Dp, 85.75.-d
\item[DOI] 10.12693/APhysPolA.volume.page
\end{description}
\end{abstract}

\maketitle
\section{Introduction}
The generation of a transverse spin polarization by an electric field (Edelstein Effect) and its inverse (Inverse Edelstein Effect) have attracted much interest from both theoretical and experimental points of view in recent years, thanks to the potential for spintronics applications of these effects. A recent review can be found in Ref.[\onlinecite{ganichev2016spin}].
The microscopic origin of the effect lies in the spin-orbit interaction (SOI).
Usually, SOIs are classified as intrinsic when due to the  structure inversion asymmetry (Rashba\cite{bychkov1984properties}) and/or bulk inversion asymmetry (Dresselhaus\cite{dresselhaus1955spin}), whereas  extrinsic ones are due to random scattering from impurities. The interplay of intrinsic and extrinsic SOIs in the current-induced spin polarization (CISP) was considered in Ref.[\onlinecite{raimondi2011spin}] where only the Rashba SOI (RSOI) was taken into account.
There it was shown that the interplay depends on the ratio of the two main spin relaxation mechanisms active in a two-dimensional electron gas (2DEG). Spin relaxation due to SOI from impurities is usually referred to as the  Elliott-Yafet (EY) mechanism  and in this case the spin relaxation time scales as the momentum relaxation time. Intrinsic SOI yields in addition the D'yakonov-Perel' (DP) spin relaxation due to the precessional mechanism, where the spin relaxation time scales as the inverse of the momentum relaxation time.
It was pointed out \cite{giglberger2007rashba,ganichev2004experimental} that  the CISP in semiconductors can be strongly anisotropic due to the  interplay of RSOI  and DSOI  in the presence of impurity scattering\cite{trushin2007anisotropic}.  As noted in Ref.[\onlinecite{ganichev2016spin}], the anisotropy of the spin accumulation may be 
exploited  for spin field transistors operating in the non ballistic regime\cite{schliemann2003nonballistic}.
It is then relevant to extend the results of Ref.[\onlinecite{raimondi2011spin}] to the case when both RSOI and DSOI, as well as SOI from impurities,  are present. This is the aim of the present paper. Moreover, in contrast to what was done in Ref.[\onlinecite{raimondi2011spin}], where the quasiclassical Keldysh Green function technique was used, we adopt here the diagrammatic language and the Kubo formula, which allows to identify the different physical contributions to the Edelstein effect or CISP.
We will show in particular that the contributions due to RSOI and DSOI can cancel each other for equal RSOI and DSOI strengths. 

The layout of the paper is the following. In the next section we introduce the Kubo formula and evaluate the Edelstein effect arising from the intrinsic SOI.
In section III we will evaluate side-jump and skew-scattering contributions  to the Edelstein current-induced polarization. 
A brief conclusion is provided in section IV.  
   
\section{Linear response theory}
The model Hamiltonian for a two-dimensional electron gas (2DEG)  in the presence of the spin-orbit interaction reads
\begin{eqnarray}
H&=&\frac{p^2}{2m}+\alpha (p_y\sigma_x- p_x\sigma_y)+\beta( p_x\sigma_x-p_y\sigma_y) \nonumber\\
&-&\frac{\lambda_0^2}{4}\sigmabold\times \nabla V(\rv)\cdot \pv+V(\rv),\label{ham}
\end{eqnarray}
with $\pv=-i\hbar\nabla_{\rv}$ the momentum operator and $V(\rv)$ representing a short-range impurity potential. In Eq.(\ref{ham}) $m$ is the effective mass in the sample, ${\boldsymbol\sigma}=(\sigma_x,\sigma_y,\sigma_z)$ the vector of Pauli matrices, $\alpha$ and $\beta$ the Rashba and Dressehaus spin-orbit coupling constants. Finally $\lambda_0$ is the effective Compton wavelength. 
We assume the standard model of white-noise disorder potential 
 with $\langle V(\rv)\rangle=0$
 and $\langle V(\rv_1)V(\rv_2)\rangle =n_iv_0^2\delta(\rv_1-\rv_2)$, $n_i$ being the impurity concentration.  
 In the following, for the treatment of the  skew-scattering effect, we will also need the third moment of the disorder distribution $\langle{V(\rv_1)V(\rv_2)V(\rv_3)}\rangle=n_i v_0^3 \delta(\rv_1-\rv_2)\delta(\rv_2-\rv_3)$. 
 In the following, for the sake of simplicity, we choose units such that $\hbar=1$.
 
\begin{figure}
\centering
\includegraphics*[width=0.8\linewidth]{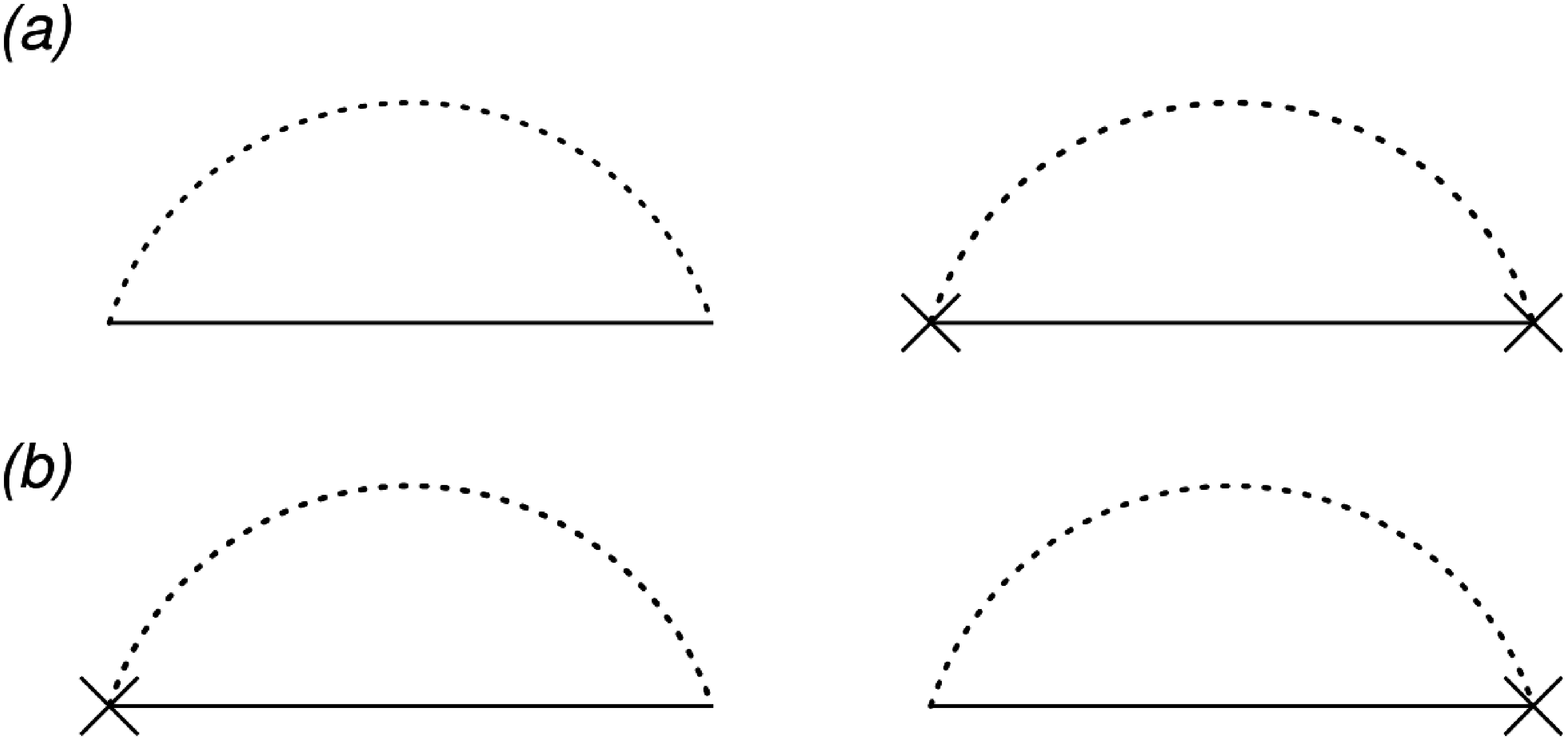}
\caption{Lowest order diagrams for the disorder-averaged selfenergy.  (a) The self-consistent Born approximation for the  spin-independent ($\Sigma_0$) and spin-dependent ($\Sigma_{EY}$) selfenergies. (b) The extra correction to the selfenergy due to the interplay of RSOI and extrinsic SOI.
 The dashed line denotes the impurity average and the cross denotes the spin-orbit insertion from the impurity potential. }\label{self energy}
\end{figure}

In linear response theory, the spin polarization along the y direction 
due to an electric field  applied along the x direction is given by 
\begin{equation}
\label{LRT}
S^y=\sigma^{yx}_{EC}E_x
\end{equation}
where $\sigma^{yx}_{EC}$ is the Edelstein conductivity \cite{edelstein1990solid} given by the Kubo formula\cite{shen2014microscopic}
\begin{equation}
\sigma^{yx}_{EC}=\frac{(-e)}{2\pi} \sum_\pv{\rm Tr} [G^A\frac{\Gamma_y}{2}G^RJ_x],
\label{Kuboo}
\end{equation}
where $\Gamma_y$ is the  spin vertex renormalized by impurity scattering and $J_x$ is the number current vertex. In the presence of RSOI and DSOI, the retarded Green function has a structure in spin space, which can be expanded in Pauli matrix basis in the form 
\begin{equation}
G_{\pv}^{R}=G^{R}_0\sigma_0 +G^{R}_x\sigma_x+G^{R}_y\sigma_y
\label{greens function}
\end{equation}
where 
\begin{eqnarray}
G^{R}_0&=&\frac{G^{R}_++G^{R}_-}{2} \nonumber\\
G^{R}_x&=&(\alpha\hat{p}_y +\beta \hat{p}_x)\frac{G^{R}_+-G^{R}_-}{2\gamma}\nonumber\\
G^{R}_y&=&-(\alpha \hat{p}_x +\beta \hat{p}_y)\frac{G^{R}_+-G^{R}_-}{2\gamma}.
\end{eqnarray}
with $G^R_{\pm }=(\epsilon-\frac{p^2}{2m} \mp \gamma p+\frac{i}{2\tau_\pm})^{-1}$, and the advanced Green function is obtained via the relation $G^A_{\pm}=(G^R_{\pm})^*$. $\gamma^2=\alpha^2+\beta^2+2\alpha \beta (\hat{p}_x\hat{p}_y+\hat{p}_y\hat{p}_x)$ is the total spin-orbit strength and depends on the direction of the momentum $\hat{p}_x=\cos(\phi)$ and $\hat{p}_y=\sin(\phi)$. 
Within the self-consistent Born approximation the selfenergy is given by the diagrams of Fig.\ref{self energy}(a) and has two contributions due to spin-independent and spin-dependent scatterings\cite{edelstein1990solid,shen2014theory}
 \begin{eqnarray}
 \Sigma^R &=&\Sigma^R_0+\Sigma^R_{EY} = n_iv_0^2\sum_{\pv'}G^R_{\pv'}\nonumber\\
 &+& n_iv_0^2(\frac{\lambda_0^2}{4})^2\sum_{\pv'}\sigma^z G^R_{\pv'}\sigma^z(\pv\times \pv')^2_z
 \nonumber\\
 &=&-{\rm i}\frac{1}{2\tau_0}-{\rm i}\frac{1}{4\tau_{EY}}=-{\rm i}\frac{1}{2\tau}, 
 \label{selfenergy}
 \end{eqnarray} 
where ${1}/{2\tau}$ is the total quasiparticle relaxation rate. Whereas the first term, to zero order in $\lambda^2_0$, yields the standard elastic scattering time,  the second one, to second order in $\lambda^2_0$,  is responsible for the EY spin relaxation. The standard expression for the spin-independent scattering and EY spin relaxation rates is given by
\begin{equation}
\frac{1}{\tau_0}=2\pi n_iN_0\upsilon^2, \   \frac{1}{\tau_{EY}}=\frac{1}{\tau_0}\left(\frac{\lambda_0 p_F}{2}\right)^4,
\end{equation} 
where $N_0=m/(2\pi)$ and $p_F$ are   the density of states and the Fermi momentum, respectively,  of the 2DEG in the absence of spin-orbit coupling.

In order to evaluate Eq.(\ref{Kuboo}) we have introduced the matrix element of the number current vertex $J_{x}$ from state $\pv'$ to state $\pv$ 
 \begin{eqnarray}
 J_{x,\pv\pv'}=\delta_{\pv\pv'}\left( \frac{p_x}{m}-\alpha \sigma_y+\beta\sigma_x\right)+\delta J_{x,\pv\pv'}.\label{Jx}
 \end{eqnarray}      
The latter term $\delta J_{x,\pv\pv'}$  is responsible for the side-jump contribution to the Edelstein conductivity and will be discussed further in Section III. 

The renormalized spin vertex may be expanded   in Pauli matrices as $\Gamma_y=\Sigma_{\eta} \Gamma_y^ {\eta} \sigma^{\eta}$ and is obtained by  summing ladder diagrams. As a result the vertex obeys
an integral equation, which within the  standard approximation, becomes an algebraic one
\cite{schwab2002magnetoconductance} 
 \begin{eqnarray}
\Gamma_{y}^{\eta}&=&\delta_{y\eta}+\frac{1}{2}\sum_{\mu \upsilon i} I_{\mu \upsilon} Tr[\sigma_\eta \sigma_\mu\sigma_i\sigma_\upsilon]\Gamma^i_{y} \nonumber\\
&+&\frac{1}{2}\sum_{\mu \upsilon i} J_{\mu \upsilon} Tr[\sigma_\eta \sigma_z \sigma_\mu\sigma_i\sigma_\upsilon \sigma_z]\Gamma^i_{y},
\label{vertex}
\end{eqnarray}
where we have defined 
\begin{equation}
I_{\mu \upsilon}=\frac{1}{2\pi N_0 \tau_0}\sum_{\pv'} G^R_\mu G^A_\upsilon, \
J_{\mu \upsilon}=\frac{1}{4\pi N_0 \tau_{EY}}\sum_{\pv'} G^R_\mu G^A_\upsilon.\label{integrals}
\end{equation}   
Symmetry arguments in Eq.(\ref{vertex}) indicate that, when both Rashba and Dresselhaus are present, the renormalized spin vertex $\Gamma_y$ is not simply proportional to $\sigma_y$, but acquires components on both $\sigma_x$ and $\sigma_y$. Upon the integration over the momentum in Eq.(\ref{integrals}), 
some of the integrals $I_{\mu \upsilon}$ are zero and so the equations simplify. As a result we finally obtain
\begin{equation}
	\label{vertex_y}
	\begin{pmatrix}
	\Gamma^y_y\\
	\Gamma^x_y
	\end{pmatrix}=
		\begin{pmatrix}
			1-I_{00}+J_{00} &   -2(I_{yx}-J_{yx})\\
			-2(I_{xy}-J_{xy})        &   1-I_{00}+J_{00}
		\end{pmatrix}^{-1}
			\begin{pmatrix}
				1\\
				0
			\end{pmatrix}
\end{equation}
where
\begin{eqnarray}
	1-I_{00}+J_{00}&\simeq &\tau(\frac{1}{\tau_{\alpha}}+\frac{1}{\tau_{\beta}}+\frac{1}{\tau_{EY}})\simeq\frac{\tau}{\tau_t}\label{dia_rate},\\
	-2(I_{xy}-J_{xy})&\simeq &\frac{2\tau}{\tau_{\alpha \beta}}.\label{off_dia_rate}
\end{eqnarray}
 In the diffusive regime, $\frac{1}{\tau_{\alpha}}\backsimeq(2m\alpha)^2D$, $\frac{1}{\tau_{\beta}}\backsimeq(2m\beta)^2D$ and $\frac{1}{\tau_{\alpha\beta}}\backsimeq(2m)^2\alpha \beta D$ are the DP relaxation times due to RSOI and DSOI, respectively,  and the interplay of them.  
 
Once the renormalized spin vertex is known, the Edelstein conductivity from Eq.(\ref{Kuboo}) can be put in the form
\begin{equation}
\sigma^{yx}_{EC}=\sum_{\eta=x,y}\Gamma^\eta_y \Pi_\eta, \ 
\label{EE}
\end{equation}
where the bare "Edelstein conductivity" without the contributions of the side-jump term and skew-scattering mechanisms is given by
\begin{equation}
\Pi_\eta=\frac{(-e)}{2\pi}\sum_{\pv} {\rm Tr} [G^A\frac{\sigma^\eta}{2}G^RJ_x].
\label{pi}
\end{equation}
To derive the CISP, we rewrite Eq.(\ref{LRT}) by using Eq.(\ref{EE})
\begin{eqnarray}\label{Bloch}
S^y=
\begin{pmatrix}
\Gamma^y_y &  \Gamma^x_y\\
\end{pmatrix}
\begin{pmatrix}
\Pi_{y}\\
\Pi_{x}
\end{pmatrix}E_x.
\end{eqnarray}
By using the standard technique to evaluate the integration over the absolute value of the momentum,
 the bare conductivities in Eq.(\ref{pi})  read
\begin{eqnarray}
 \Pi_{y}= 
 \tau S_\alpha\langle\frac{1}{\tau_{\gamma}}-\frac{2}{\tau_{\gamma}}\frac{\beta^{2}}{\gamma^2}\rangle,\  \Pi_{x}=
  -\tau S_\beta\langle\frac{1}{\tau_{\gamma}}-\frac{2}{\tau_{\gamma}}\frac{\alpha^{2}}{\gamma^2}\rangle \label{bar}
 \end{eqnarray}
 where 
 \begin{eqnarray}
S_\beta&=&-eN_0\tau\beta E_x\label{Ed_ax},\     \
S_\alpha=-eN_0\tau\alpha E_x.\label{Ed_by}
\end{eqnarray}
and $\langle...\rangle$ denotes the average over the momentum directions.
Then the CISP, which is equivalent to the stationary solution of the Bloch equation,  is derived by inserting Eq.({\ref{vertex_y}}) and Eq.({\ref{bar}}) into Eq.({\ref{Bloch}})
\begin{eqnarray}
 S^y&=& \left[ \left(\frac{1}{\tau_{\alpha}}+\frac{1}{\tau_{\beta}}+\frac{1}{\tau_{EY}}\right)^2-\left(\frac{2}{\tau_{\beta\alpha}}\right)^2\right]^{-1}\label{Sy} \\
&\times&\langle S_\beta\frac{2}{\tau_{\alpha\beta}}\left(\frac{1}{\tau_{\gamma}}-\frac{2}{\tau_{\gamma}}\frac{\alpha^2}{\gamma^2}\right)+S_\alpha\frac{1}{\tau_t}\left(\frac{1}{\tau_{\gamma}}-\frac{2}{\tau_{\gamma}}\frac{\beta^2}{\gamma^2}\right)\rangle.\nonumber
\end{eqnarray}
In Eq.({\ref{Sy}}), ${1}/{\tau_{\gamma}}=(2\gamma p_F\tau)^2/2\tau$ is the total DP spin relaxation for a fixed direction of the momentum.  Hence, 
 Eq.(\ref{EE}) can be seen as the spin accumulation at fixed direction of the momentum averaged over the momentum directions.
  After taking the angular average of Eq.(\ref{Sy}) we may write the expression of the CISP component along the y direction in a form reminiscent of the stationary Bloch equation
\begin{eqnarray}
S_{int}^y&=&  \left[ \left(\frac{1}{\tau_{\alpha}}+\frac{1}{\tau_{\beta}}+\frac{1}{\tau_{EY}}\right)^2-\left(\frac{2}{\tau_{\beta\alpha}}\right)^2\right]^{-1}\nonumber \\
&\times&S_\alpha(\frac{1}{\tau_{\alpha}}-\frac{1}{\tau_{\beta}})(\frac{1}{\tau_{\alpha}}-\frac{1}{\tau_{\beta}}+\frac{1}{\tau_{EY}}).\label{Syy}
\end{eqnarray}
We have added a suffix {\it int} to remind that we are only considering the intrinsic mechanism, which can be defined as the term that survives when the extrinsic SOI $\lambda_0$ vanishes. One must however borne in mind that this intrinsic term is modified by the presence of the extrinsic SOI via the appearance of the EY spin relaxation time. The consideration of the extrinsic mechanisms, i.e. those terms which only arise when the extrinsic SOI is present, will be done 
 in the next section.
 
Eq.(\ref{Syy}) generalizes to the presence of the DSOI the expression  for the 
{\it intrinsic} contribution to the  Edelstein polarization presented in  Eq.(36) of Ref.[ \onlinecite{raimondi2011spin}] and, indeed,
 reduces to it when $\beta=0$.  Furthermore, when also $\lambda_0=0$ it 
reproduces the Edelstein result for the Rashba model\cite{edelstein1990solid}.
 We also note that, in the absence of the extrinsic SOI, when $\alpha=\beta$, the intrinsic spin-orbit interaction reduces to a pure gauge  field and as such can have no physical effect\cite{tokatly2010}. In this case indeed Eq.(\ref{Syy}) predicts that the Edelstein effect vanishes.
 
 The fact that the spin vertex $\Gamma_y$ has both $\sigma_x$ and $\sigma_y$ components implies that there will be spin polarization also along the x direction.
 By performing a similar calculation for the CISP along the x direction we find
\begin{eqnarray}
S_{int}^x&=&  \left[ \left(\frac{1}{\tau_{\alpha}}+\frac{1}{\tau_{\beta}}+\frac{1}{\tau_{EY}}\right)^2-\left(\frac{2}{\tau_{\beta\alpha}}\right)^2\right]^{-1}\nonumber \\
&\times&S_\beta(\frac{1}{\tau_{\alpha}}-\frac{1}{\tau_{\beta}})(\frac{1}{\tau_{\alpha}}-\frac{1}{\tau_{\beta}}+\frac{1}{\tau_{EY}}).\label{Sxx}
\end{eqnarray}  

\section{Side-jump  and skew-scattering contributions}
\begin{figure}[t]
 \begin{center}
 \includegraphics[width=2in]{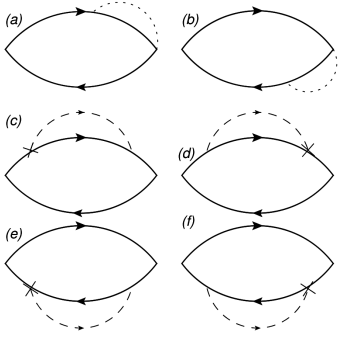}
 \caption{ Diagrams for the side-jump contribution to the Edelstein effect. The solid lines are Green's function and dashed lines represent the average over the impurity potential. The cross denotes the spin-orbit interaction from the impurity potential (a),(b) Side-jump type of diagrams originating from components proportional to $\lambda_0^2$ in the current vertices. (c)-(f) The extra corrections to the side-jump contribution due to the extrinsic effect, where the right vertex
 is for the x component of the charge current.}
 \label{diagrams_SJ}
 \end{center}
 \end{figure}

In this section we evaluate the side-jump and skew-scattering contributions to  the Edelstein conductivity. 
The selfenergies, to order $\lambda_0^2$,  in Fig.\ref{self energy}(b) are usually zero in the absence of intrinsic spin-orbit interaction due to symmetry reasons. However, when RSOI and DSOI are present, they no longer vanish and, actually, their contribution is crucial to get the full side-jump contribution to the Edelstein conductivity. Hence, the diagrams we need to consider for the side-jump mechanism are those depicted in Fig.(\ref{diagrams_SJ}). 
Diagrams shown in Figs.\ref{diagrams_SJ}(a) and \ref{diagrams_SJ}(b) correspond to the ordinary side-jump diagrams as those used to evaluate the spin Hall conductivity and originate from the anomalous correction to the current vertex to order $\lambda_0^2$.
The other diagrams shown in Figs.\ref{diagrams_SJ}(c-f)  take into account the selfenergy corrections mentioned above.
 To keep the discussion as simple as possible, we confine first to the case when only RSOI is present. The extension to the DSOI is straightforward. 
  
The anomalous current vertex from state $\pv$ to state $\pv'$ can be put in the form
 \begin{eqnarray}
 \delta J_{\pv,\pv'}^x={\rm i}\frac{v_0\lambda_0^2}{4} (p_y-p_y')\sigma_z.\label{currentVertex}
 \end{eqnarray}

 By replacing the spin current $J_{x\pv\pv'}$ in Eq.({\ref{pi}}) by $\delta J_{x\pv\pv'}$, the diagrams in Figs.\ref{diagrams_SJ}(a)  and \ref{diagrams_SJ}(b) read
\begin{eqnarray}
\Pi^{sj(a+b)}_{y}&=&{\rm -i}\frac{e v_0^2 n_i }{2\pi}\frac{\lambda_0^2}{4}\sum_{\pv\pv'}(p_y'-p_y)
\nn\\
&\times&\frac{1}{2}{\rm Tr} \left[ G^A_{\pv}\sigma_yG^R_{\pv}(G^R_{\pv'}\sigma_z-\sigma_zG^A_{\pv'})\right].
\label{sj_diagram}
\end{eqnarray}
The diagrams in Figs.\ref{diagrams_SJ}(c) and \ref{diagrams_SJ}(f) corresponding to the contributions from the selfenergy renormalization of the Green functions are given by
\begin{eqnarray}
\Pi^{sj(c+d)}_{y}&=&{\rm i}\frac{e  n_i }{2\pi}v_0^2\frac{\lambda_0^2}{4}\sum_{\pv\pv'}
{\rm Tr}\left[\frac{\sigma_y}{2}G^R_{\pv}[G^R_{\pv'}({\pv}'\times{\pv})_z\sigma_z\right.\nn\\
&+&\left.  ({\pv}\times{\pv}')_z\sigma_zG^R_{\pv'}]G^R_{\pv}\frac{{p}_x}{m}G^A_{\pv}\right],
\label{sj_diagram34}
\end{eqnarray}
\begin{eqnarray}
\Pi^{sj(e+f)}_{y}&=&{\rm i}\frac{e  n_i }{2\pi}v_0^2\frac{\lambda_0^2}{4}\sum_{\pv\pv'}
{\rm Tr}\left[\frac{\sigma_y}{2}G^R_{\pv}\frac{{p}_x}{m}G^A_{\pv}\right.\\
&\times &\left. [({\pv}\times{\pv}')_z\sigma_zG^A_{\pv'}\nn+G^A_{\pv'}({\pv}'\times{\pv})_z\sigma_z]G^A_{\pv}\right].
\label{sj_diagram56}
\end{eqnarray}
After performing the integration over the momentum $\pv'$ and using the expansion of the Green function in Pauli matrices, we obtain 
\begin{eqnarray}
\label{sj_3}
\Pi^{sj(a+b)}_{y} &=&{\rm i}\frac{e}{4\tau_0}\frac{\lambda_0^2}{4}\frac{1}{2\pi}\sum_{\pv}p
\left( G^A_{+}G^R_{-}-G^A_{-}G^R_{+}\right)\nn\\
&=&\frac{\lambda_0^2p_F^2}{4}S_0
\end{eqnarray}
with $S_0=-eN_0 \tau$ and
\begin{eqnarray}
\label{sj_3456}
\Pi^{sj(c+d+e+f)}_{y} &=& (4)(-i)\frac{e}{4\tau_0}\frac{\lambda_0^2}{4}\frac{\alpha}{2\pi}\sum_{\pv}p_x^2\\
&\times&{\rm Tr}[\sigma_yG^R_{\pv}\sigma_yG^R_{\pv}G^A_{\pv}]=\frac{\lambda_0^2p_F^2}{4}S_0.\nn
\end{eqnarray}
By collecting the result of all the diagrams, one gets
\begin{equation}
\label{sj_abcdef}
\Pi^{sj}_{y}=\Pi^{sj(a+b)}_{y}+\Pi^{sj(c+d+e+f)}_{y}=2\frac{\lambda_0^2p_F^2}{4}S_0. 
\end{equation}
Then, recalling that the side-jump spin Hall conductivity reads
\begin{equation}
\label{sj_sh}
\sigma^{SHE}_{sj}=-\frac{e}{2\pi} \frac{\lambda_0^2p_F^2}{4},
\end{equation}
\begin{figure}[t!]
\begin{center}
\includegraphics[width=2in]{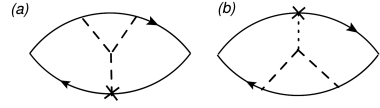}
\caption{Diagrams for the skew-scattering contribution to the Edelstein effect. The cross
denotes the correction to the Green function due to SO scattering.}
\label{diagrams_SS}
\end{center}
\end{figure}
we finally obtain
\begin{equation}
\label{sj_4}
\sigma^{sj}_{EC,yx}=-2\tau_s m \alpha \sigma^{SHE}_{sj}.
\end{equation}
The above term will then give the following contribution to the CISP 
\begin{equation}
\label{sj_5}
S^y= -2m\alpha \tau_{s} \sigma^{SHE}_{sj}E_x,
\end{equation}
with the total relaxation rate being $\frac{1}{\tau_s}=\frac{1}{\tau_{EY}}+\frac{1}{\tau_{\alpha}}$. Note that by identifying the side-jump contribution to the spin Hall current as $J_y^z=\sigma^{SHE}_{sj}E_x$, one obtains the same expression as in Ref.[\onlinecite{raimondi2011spin}] as expected in the Bloch equation for $S^y$ when extrinsic contributions are explicitly taken into account.

Finally, we  proceed to evaluate the diagrams responsible for the skew-scattering contribution to the bare conductivity. The diagrams in Fig.\ref{diagrams_SS} give 
 \begin{eqnarray}
 \label{ss_1}
 \Pi^{ss(a+b)}_{y} &=&-i\frac{e}{2\pi}v_0^3\frac{\lambda_0^2}{4}\sum_{\pv \pv'\pv''}Tr\left[\frac{\sigma_y}{2}G^R_{\pv}[G^R_{\pv'}\right.\nn\\
 &\times&\left. G^R_{\pv''}\frac{p_x''}{m}G^A_{\pv''}({\pv}''\times{\pv})_z\sigma_z\right.\nn\\
 &+&({\pv}\times{\pv}')_z\sigma_zG^R_{\pv'}\frac{p_x'}{m}G^A_{\pv'}G^A_{\pv''}]G^A_{\pv}].\nn\\
 \end{eqnarray}
 Similarly to Eqs. (\ref{sj_diagram34}) and (\ref{sj_diagram56}), after taking the integration over $\pv'$ and $\pv''$ and using the expansion of the Green function, we can obtain
  \begin{eqnarray}
   \label{ss_2}
   \Pi^{ss(a+b)}_{yx} = i\frac{e v_0p_F ^2}{4m}N_0\frac{\lambda_0^2}{4}\sum_{\pv}p\frac{1}{2}( G^R_{-}G^A_{+}-G^A_{+}G^R_{-}).\nn\\
   \end{eqnarray}
 Finally the total skew-scattering contribution for a screened impurity potential gives 
 \begin{equation}
 \label{ss_5}
 S^y= -2m\alpha \tau_{s}  \sigma^{SHE}_{ss}E_x.
  \end{equation} 
  where $\sigma^{SHE}_{ss}$ is the spin Hall conductivity associated to the  skew-scattering mechanism. 
Similar to the side jump, the skew-scattering contribution can be included in the Edelstein conductivity, which amounts to say that  $\sigma^{yx}_{EC,sj}$ can be replaced by the sum of  both contributions $\sigma^{yx}_{EC,sj}\rightarrow \sigma^{yx}_{EC,sj}+ \sigma^{yx}_{EC,ss}$. 

The inclusion of the DSOI is straightforward, although the calculation is lengthy. However, the final result can be guessed by carefully considering the results
(\ref{Syy}) and (\ref{Sxx}).  In Eq.(\ref{Syy}), for instance, one sees that the spin-orbit interaction determines the form of the spin polarization in three respects. First,  there is a factor $S_\alpha$ reminiscent of the Edelstein effect in the RSOI only model. Secondly, the DSOI only appears in the specific element of the inverse matrix of the scattering rates. Finally, the factor $1/\tau_\alpha -1/\tau_\beta$ can be interpreted as due to the intrinsic spin Hall conductivity $\sigma^{SHE}_{int}=(e/8\pi)(2\tau/\tau_\alpha -2\tau/\tau_\beta)$. 
For $S^x$ in Eq.(\ref{Sxx}) there is a similar situation with the roles of RSOI and DSOI interchanged.
Then, in order to have the side-jump and skew-scattering contributions to the Edelstein conductivity it is sufficient to replace the intrinsic spin Hall conductivity with $\sigma^{SHE}_{ext}=\sigma^{SHE}_{sj}+\sigma^{SHE}_{ss}$ to read
\begin{eqnarray}
S_{ext}^y&=&  \left[ \left(\frac{1}{\tau_{\alpha}}+\frac{1}{\tau_{\beta}}+\frac{1}{\tau_{EY}}\right)^2-\left(\frac{2}{\tau_{\beta\alpha}}\right)^2\right]^{-1}\nonumber \\
&\times&S_\alpha (\frac{1}{\tau_{\alpha}}-\frac{1}{\tau_{\beta}}+\frac{1}{\tau_{EY}})\frac{4\pi}{e\tau}\sigma^{SHE}_{ext}\label{Syy_sjss}
\end{eqnarray}
and
\begin{eqnarray}
S_{ext}^x&=&  \left[ \left(\frac{1}{\tau_{\alpha}}+\frac{1}{\tau_{\beta}}+\frac{1}{\tau_{EY}}\right)^2-\left(\frac{2}{\tau_{\beta\alpha}}\right)^2\right]^{-1}\nonumber \\
&\times&S_\beta (\frac{1}{\tau_{\alpha}}-\frac{1}{\tau_{\beta}}+\frac{1}{\tau_{EY}})\frac{4\pi}{e\tau}\sigma^{SHE}_{ext}.\label{Sxx_sjss}
\end{eqnarray}

The above result has been confirmed by an explicit calculation. 
 The sum of Eqs.(\ref{Syy}) and (\ref{Syy_sjss}) gives the total expression for the Edelstein polarization along the y direction. Similarly Eqs.(\ref{Sxx}) and (\ref{Sxx_sjss})
provide the corresponding expression for the polarization along the x direction.
The four equations represent then the main result of this paper.
One interesting consequence of these equations is that, by invoking the Onsager reciprocity, along, say the y, direction, should in principle yield a charge current both along the x and y directions, an effect which can be tested experimentally.

\section{Conclusions}
In summary, we have obtained an analytical formula of the Edelstein conductivity in the presence of both extrinsic and intrinsic spin orbit interaction as well as scattering from impurities. The formula is valid at the level of the Born approximation and to first order beyond the Born approximation and was obtained by standard diagrammatic techniques, then complementing the analysis of Ref.[\onlinecite{raimondi2011spin}], derived via the quasiclassical Keldysh Green function technique.
It has been shown that the current induced spin polarization can be anisotropic due to interplay of Rashba and Dresselhaus spin orbit interactions in the 2DEG. 
We also find that the interplay of intrinsic and extrinsic spin-orbit interactions may tune the value of the current-induced spin polarization depending on the ratio of the DP and EY spin relaxation rates.

%

\end{document}